# Rhythmic Motion of a Droplet under a DC Electric Field

M. Hase , S. N. Watanabe and K. Yoshikawa[*]

*Department of Physics, Graduate School of Science, Kyoto University, Kyoto 606-8502, Japan*

**Abstract**

The effect of a stationary electric field on a water droplet with a diameter of several tens micrometers in oil was examined. Such a droplet exhibits repetitive translational motion between the electrodes in a spontaneous manner. The state diagram of this oscillatory motion was deduced; at 0-20 V the droplet is fixed at the surface of the electrode, at 20-70 V the droplet exhibits small-amplitude oscillatory motion between the electrodes, and at 70-100 V the droplet shows large-amplitude periodic motion between the electrodes. The observed rhythmic motion is explained in a semi-quantitative manner by using differential equations, which includes the effect of charging the droplet under an electric field. We also found that twin droplets exhibit synchronized rhythmic motion between the electrodes.

**Text body**

In chemistry and chemical engineering, the effect of an electric field on a multiphase system such as an emulsion has attracted increasing attention due to the interesting behaviors that have been observed; e.g. deformation and breakup of a single droplet [1-3], formation of a fiber-like chain of droplets [4], return motion [2] and so on. Active studies on the behavior of multiphase flow under a direct electrostatic field have been performed using numerical simulations and experimental methods. However, the mechanism is still poorly understood due to its complexity, in that the effects of inertia, viscosity, surface deformation, and electrohydrodynamic stresses all interact.

On the other hand, among various kinds of microemulsions, water droplets with a diameter of several tens of microns have recently attracted attention as a convenient model of living cells [5-7]. Active studies on manipulation techniques have been performed using laser tweezer [8] and micro fluidics [9], where micro droplets simplify the physical assessment of fluid behavior compared to centimeter-sized droplets because at a micrometer scale the effect of inertia on the behavior of the fluid is less important than those of surface tension and viscosity. To examine the effect of an electric field on micro droplets for the purpose of manipulation, we directly observe the repetitive translational motion of micro droplets between two electrodes under the application of DC voltage, which reflect the presence of limit-cycle oscillation in real space.

A surfactant (DOPC, phospholipid with a neutral head group, purchased from WAKO) was solved in oil (rapeseed oil purchased from NAKARAI) by sonication for 90 min and used within 24 hours. To prepare micro-water droplets, 2 μl water was added to 200 μl oil containing 10 μM surfactant. Droplets with a diameter of several tens of μm were obtained after several seconds of vortex agitation.

A single droplet was situated between a pair of gold rods as electrodes with a diameter of 100 μm and separated by about 400 μm, and then DC voltage, $\psi = 0\text{-}100V$, was applied, where the droplet is charged at each electrode (Fig.1A (a)). We set x-axis on the coaxial line of the electrodes and the y-axis perpendicular to the x-axis, where origin is midway between the electrodes (Fig. 1A (b)). We used a phase-contrast microscope (Nikon TE-300) equipped with a micromanipulation system (Narishige) and a CCD camera that recorded 30 flames per second.

Figure 1B shows microscopic images of a single droplet between the electrodes at 100 V. The droplet exhibits repetitive motion without deformation or breaking up. Figure 1C shows a plot of position of the droplet (x-value upper panel; y-value lower panel). The droplet moves along the coaxial line of the electrodes.

This oscillatory motion is also generated regardless of the charge and the type of the surfactant, for example Tween 20 (neutrally charged surfactant) and DOPG (phospholipid with a negatively charged head group). Small air bubbles are found to be generated within the droplet undergoing rhythmic motion, which indicates that the droplet is charged by the electrolysis of water. The gradient of the applied electronic field is greatest along the line between the tips of the electrodes, indicating that attractive force on the droplet as a dielectric body acts in the direction of $y = 0$.

When the voltage is well below 20 V, a droplet is pulled toward either electrode and is attached to the electrode surface, as exemplified in Fig. 2A(a) (10 V). At high electrical potentials, "small oscillation" of a droplet is generated between the origin and an electrode, as exemplified in Fig. 2A(b) (50 V), while small oscillation is observed around each electrode. When the voltage exceeds 70 V, "large oscillation" is generated between the electrodes, where the droplet touches both electrodes, as show in Fig. 2A(c) (100 V). The trajectory of droplet motion is on the coaxial line between the electrodes in both small and large oscillations.

At a micrometer scale, viscosity and surface tension should strongly influence fluid behavior. To evaluate the most important effect under the conditions in the study, the capillary number, Ca, for comparing viscosity and surface tension and the Reynolds

number, Re, for comparing inertia and viscosity are written, respectively, as

$$Ca = \frac{\eta v}{\gamma} \sim 10^{-2} \quad (1)$$

$$Re = \frac{\rho v r}{\eta} \sim 10^{-4} \quad (2)$$

where $\eta$ is the viscosity of the oil ($\approx 8.4 \times 10^{-2} [Pa \cdot s]$), $v$ is the translational velocity of a droplet ($\sim 10^{-3} [m/s]$), $\gamma$ is the surface tension between water and oil ($\sim 10^{-2} [N/m]$), $\rho$ is the density of the oil ($\approx 0.91 \times 10^{3} [kg/m^3]$) and $r$ is the size of a droplet ($\sim 10^{-5} [m]$). Next, the electric capillary number, Ce, for comparing electrostatic forces and surface tension, is written as

$$Ce = \frac{\varepsilon_w \varepsilon_0 E^2 r}{\gamma} \sim 10^{-1} \quad (3)$$

where $\varepsilon_w$ is the dielectric constant of water, $\varepsilon_0$ is the permittivity of vacuum, $E$ is the electrostatic field around a droplet ($E \approx \psi/L$, $L$ is the distance between the electrodes). These considerations suggest that the deformation of a droplet from a spherical shape is negligible and that the effect of inertia on motion is also minimal.

In the stationary state, the droplet wets an electrode. It is expected that a droplet would move away form an electrode when repulsive electrostatic force overcomes the attraction of the droplet to the surface of the electrode due to the surface tension, which is written as

$$\gamma l < qE \quad (4)$$

where $l$ is the circumference of the wetting area and $q$ is the charge of the droplet.

The dynamic equation for the position of a droplet can be written as

$$k\dot{x} = qE - \nabla U(x) \quad (5)$$

where we ignored the effect of inertia, $k$ is a coefficient of resistance ($k = 6\pi\eta r$), and $U(x)$ is potential. In the experiment, an uncharged droplet is attracted to the nearest electrode regardless of whether the voltage is positive or negative. By considering the attractive force between a droplet and each electrode due to the induced charge of the droplet, by symmetry, $U(x)$ can be written as,

$$U(x) = a\psi^2 x^2 \quad (6)$$

where $a$ is a negative constant, we consider that the potential due to the induced charge is proportional to $\psi^2$ and we ignore terms of more than third order of x for simplicity. By taking the effect of charge leakage into consideration, the time-dependent change in the charge may be given as,

$$\dot{q} = -\alpha q \quad (7)$$

where $\alpha$ is a positive constant. We consider that a droplet gets a charge $q = \pm C\psi$ when in contact with an electrode ($C = 2\pi^3 r^2 \varepsilon\varepsilon_0 / 3L$, $C$ is capacitance, where $\varepsilon$ is the dielectric constant of oil ($\approx 5$). [10]). We set new parameters $X$ and $Q$ as $x = \frac{L}{2}X$ and $q = CQ$. With a linear stability analysis, we found that (5) and (7) have an unstable solution, $(X,Q)=(0,0)$. For simplicity $a$, $\alpha$, $r$, $L$, and $l$ are set as $-\frac{2a}{k}=2.5\times10^{-4}$, $\alpha=1$, $r=4\times10^{-5}[m]$, $L=4\times10^{-4}[m]$ and $l=2\times10^{-7}[m]$. Figure 3A (a) shows the results of a simulation at 10 V, where $(X,Q)=(\pm0.7,\pm10)$ at t=0. After a droplet touches each electrode, it remains stationary at the electrode. Figure 3A (b) shows the results of a simulation at 50 V, where $(X,Q)=(\pm0.5,\pm20)$ at t=0. A droplet exhibits periodic motion around each electrode with small amplitude, and two limit cycles exist. Figure 3A (c) shows the results of a simulation at 100 V, where $(X,Q)=(0.5,40)$ at t=0. A droplet exhibits stable oscillatory motion between the electrodes and touches both electrodes with large amplitude. Figure 3B shows a plot of the amplitude of oscillatory motion. A droplet exhibits a stationary mode below 20 V and two types of modes of repetitive motion, small and large oscillations, exist at 20-70 V and 70-100 V, respectively, where in the transition between the modes the amplitude changes in a discrete manner and the transition between small and large oscillations occurs with spontaneous spatial symmetry breaking.

We also found that twin droplets simultaneously exhibit rhythmic motion without coalescence along the coaxial line between the electrodes at 100 V as exemplified in Fig.4, suggesting that the droplets exchange charge when they touch each other. It is noted that, in Fig.4, these two oscillatory motions synchronize each other in the mode of in-phase, where the amplitudes are different each other.

In the experiments, the droplets would scarcely fuse together probably due to the charge of the droplets. The droplets convey electrons by engaging in rhythmic motion between the electrodes and by exchanging the charge when they touch each other, where the charge exchanged is determined in accordance with the Laplace Equation with boundary conditions. The orbits of the droplets are confined to the coaxial line for the same reason as in the case of a single droplet. In addition, more than two droplets, to a tested maximum of five droplets, exhibit synchronized rhythmic motion in a similar fashion. It is difficult to precisely predict the motion of multiple droplets that exhibit rhythmic motion due to their mutual electric interaction.

We have shown that periodic oscillatory motion is generated for micro water droplets in oil under a DC electronic field. The droplets are confined to the coaxial line between the tips of the electrodes, which suggests that this experimental system may be applicable as a prospective method for manipulation. The novel oscillation phenomenon observed in the present study may be associated with some aspects in the dynamic behavior in living cells, where thermodynamically open conditions are essential.

This work was supported by the JST ICORP project, by a Grant-in-Aid for the 21st Century COE "Center for Diversity and Universality in Physics", and a

Grant-in-Aid for Scientific Research on Priority Areas (No.17076007) "System Cell Engineering by Multi-scale Manipulation" from the Ministry of Education, Culture, Sports, Science and Technology of Japan. M. Hase received financial support from the Kyoto University Venture Business Laboratory.

# References

Electronic address: yoshikaw@scphys.kyoto-u.ac.jp

**Figure Captions**

**Fig. 1 A:** Schematic representation of the effect of a DC electric field on a droplet (a) and on the geometrical arrangement (b). **B:** Repetitive translational motion of a water droplet at 100V. **C:** Time traces under 100V. (a) x coordinate (b) y coordinate

**Fig. 2 A**: Spatio-temporal pictures of the motion of a droplet at 10, 50 and 100 V, respectively. **B:** Phase diagram of the oscillatory motion of a droplet as a function of the applied voltage.

**Fig. 3 A:** Numerical simulation of the motion of a droplet under a DC electric field. (a) $\psi = 10$ V with an initial condition of $(X,Q) = (\pm 0.7, \pm 10)$ symbolized as (i) and (ii), respectively. (b) $\psi = 50$ V with an initial condition of $(X,Q) = (\pm 0.5, \pm 20)$ symbolized as (i) and (ii), respectively. (c) $\psi = 100$ V with an initial condition of $(X,Q) = (0.5, 40)$ **B:** A plot of amplitude of the oscillatory motion of a droplet at different voltages.

**Fig.** 4: Spatio-temporal pictures of the synchronized rhythmic motion of twin droplets without coalescence at 100 V.

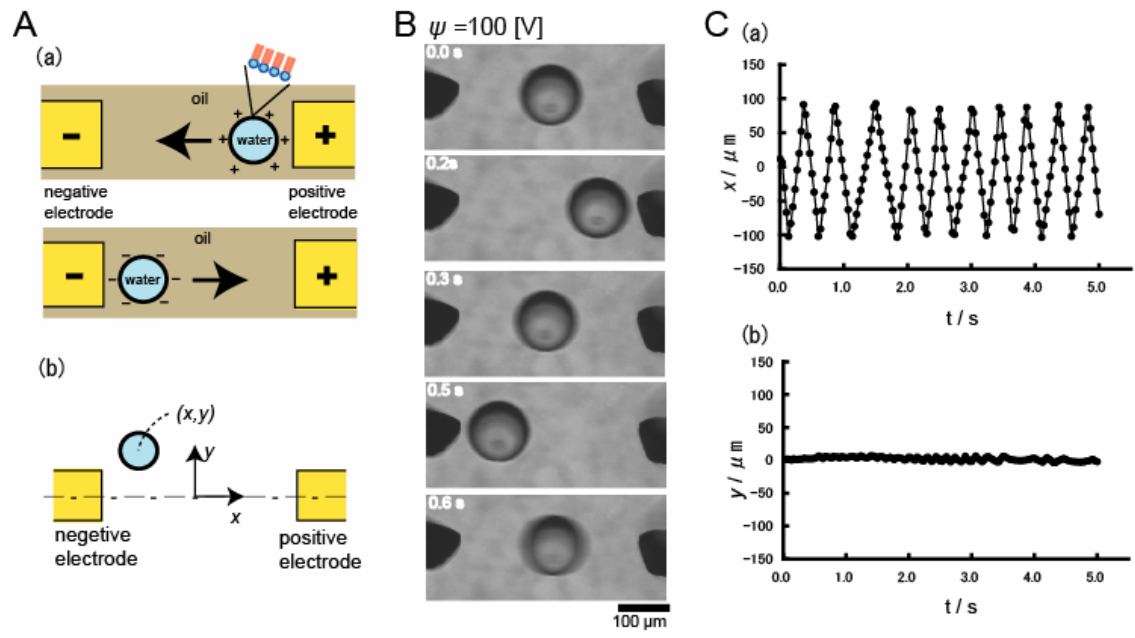

**Figure 1**

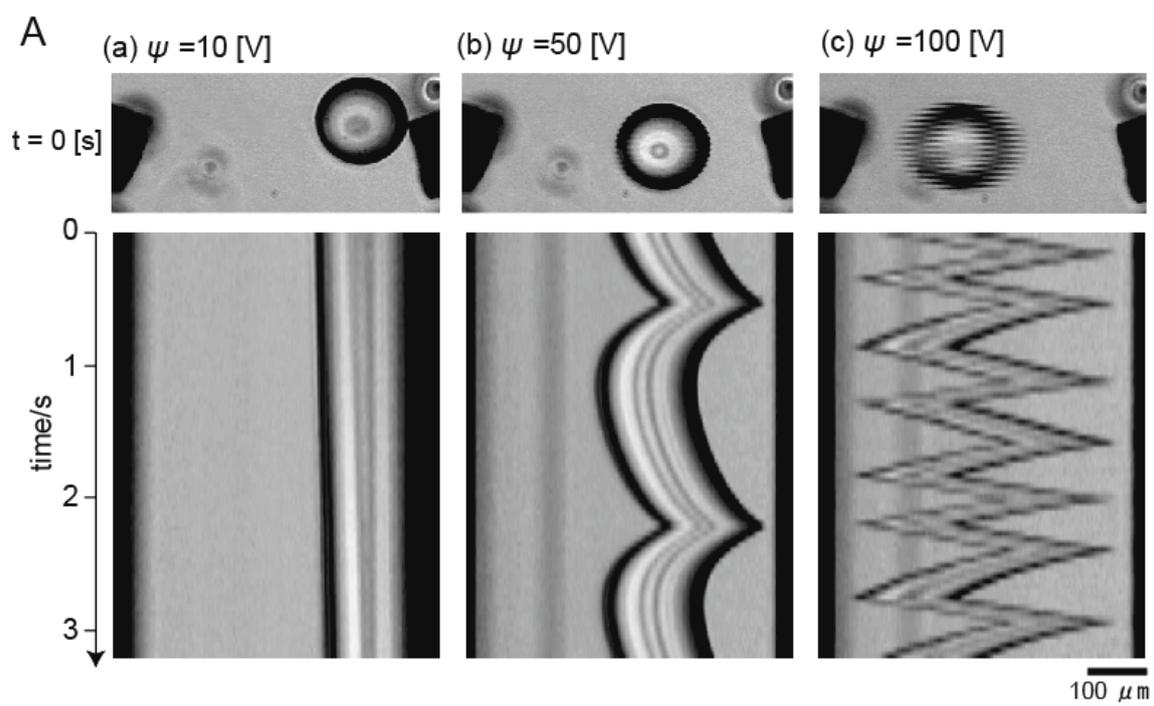

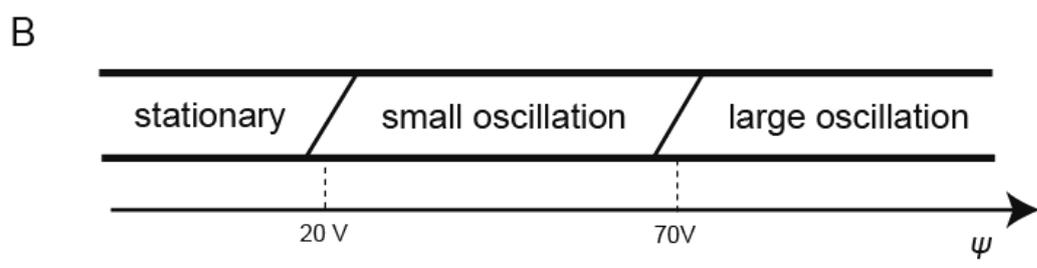

**Figure 2**

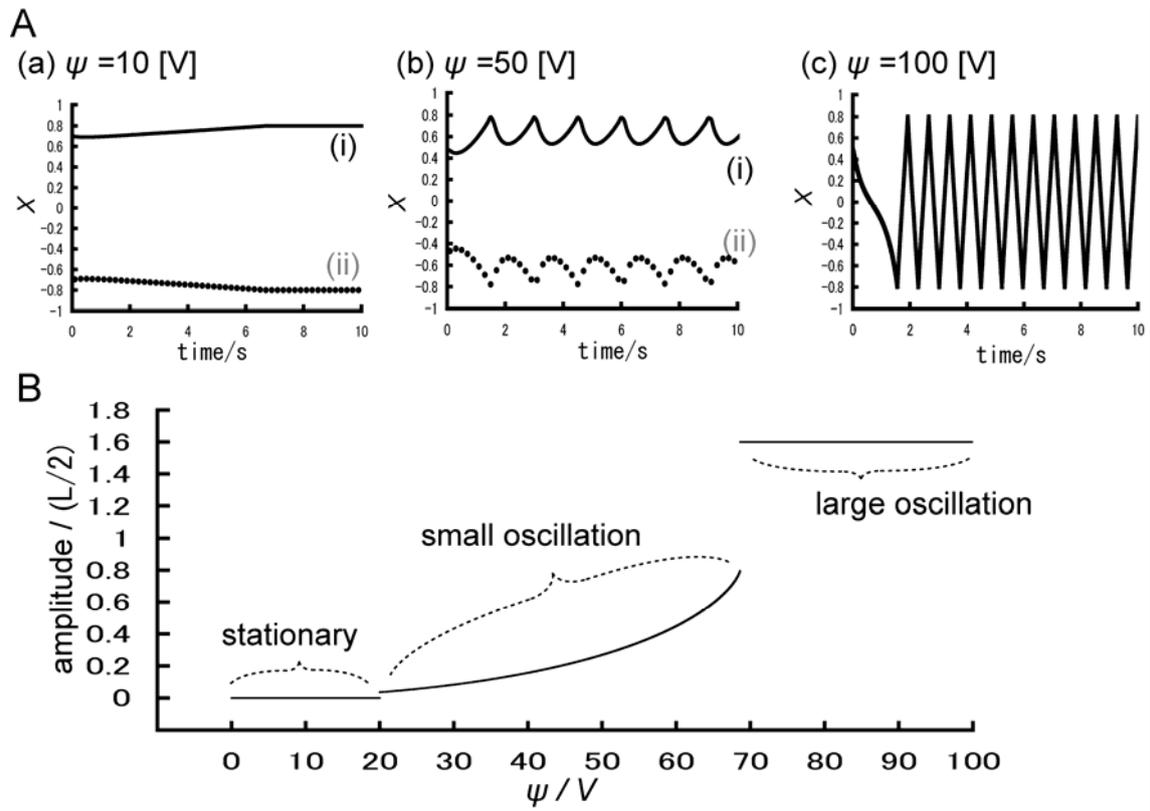

**Figure 3**

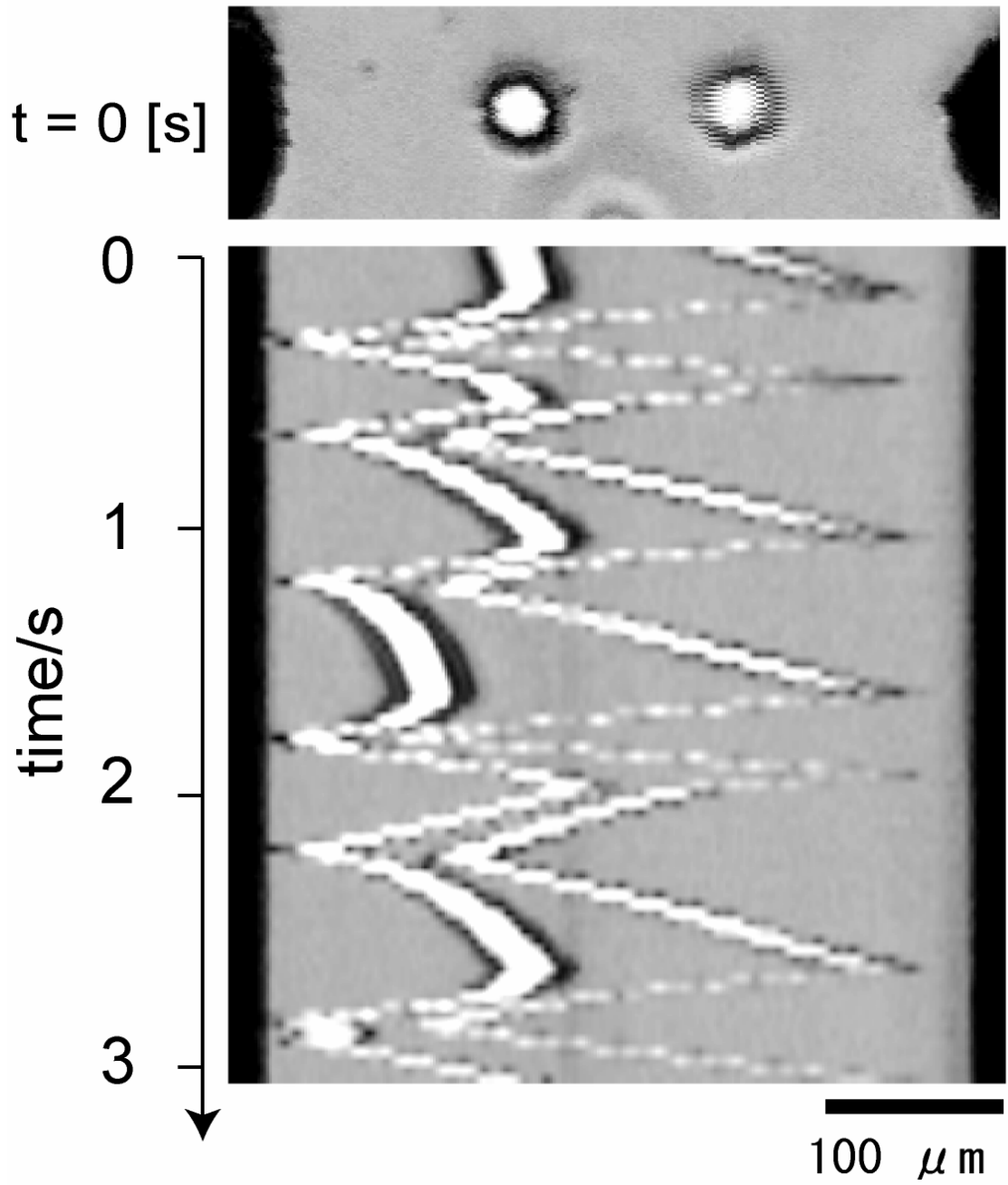

**Figure 4**